\documentclass[conference]{IEEEtran}
\IEEEoverridecommandlockouts
\usepackage{cite}
\usepackage{amsmath,amssymb,amsfonts}

\usepackage{graphicx}
\usepackage{tikz}
\usepackage{textcomp}
\usepackage{xcolor}
\usepackage{algpseudocode}
\usepackage{soul}
\def\BibTeX{{\rm B\kern-.05em{\sc i\kern-.025em b}\kern-.08em
    T\kern-.1667em\lower.7ex\hbox{E}\kern-.125emX}}
    
\begin{document}


\newcommand{\matr}[1]{\boldsymbol{#1}}
\newcommand{\mC}{\matr{C}}
\newcommand{\mW}{\matr{W}}
\newcommand{\tens}[1]{\boldsymbol{\mathcal{#1}}}
\newcommand{\tenselem}[1]{\mathcal{#1}}
\newcommand{\vect}[1]{\boldsymbol{#1}}
\newcommand{\eqdef}{\stackrel{def}{=}}
\newcommand{\E}[1]{\mathop{\mbox{$\mathbb{E}$}}\{#1\}} 
\newcommand{\I}{\mathbb I}
\newcommand{\RR}{\mathbb{R}}
\newtheorem{definition}{Definition}[section]
\newtheorem{theorem}{Theorem}[section]
\newtheorem{proposition}{Proposition}[section]
\newtheorem{lemma}{Lemma}[section]
\newenvironment{proof}{ \vspace{\smallskipamount}\par {\it Proof.}~}{\hfill $\Box$ \vspace{\medskipamount}\par }
\newcommand{\todo}{\color{red}}
\newcommand{\PC}{\color[rgb]{0.65,0,0.55}}
\newcommand{\OM}{\color[rgb]{0.3,0.6.,0.2}}
\newcommand{\SARA}{\color[rgb]{0.6,0.1,0.}}
\newcommand{\fin}{\color{black}}
\newcommand{\FIN}{\color{black}}

\title{Multivariate Normality test for colored data\\
\thanks{This work has been supported by the MIAI chair 
“Environmental issues underground” of Institut MIAI@Grenoble Alpes
(ANR-19-P3IA-0003).}
}

\author{\IEEEauthorblockN{El Bouch Sara}
\IEEEauthorblockA{\textit{Gipsa-Lab}\\
\textit{Univ. Grenoble Alpes}, France \\
sara.el-bouch@grenoble-inp.fr}
\and
\IEEEauthorblockN{Michel J. J. Olivier}
\IEEEauthorblockA{\textit{Gipsa-Lab} \\
\textit{Univ. Grenoble Alpes}, France\\
olivier.michel@grenoble-inp.fr}
\and
\IEEEauthorblockN{Comon Pierre}
\IEEEauthorblockA{\textit{Gipsa-Lab, CNRS,} \\
\textit{Univ. Grenoble Alpes}, France\\
pierre.comon@grenoble-inp.fr}

}

\maketitle

\begin{abstract} Performances of the Multivariate Kurtosis are investigated when applied to colored data, with or without Auto-Regressive prewhitening, and with or without projection onto a lower-dimensional random subspace. Computer experiments demonstrate the importance of taking into account the possible color of the  process in  calculating  the power of the normality test, in all the scenarios.

\end{abstract}


\section{Introduction}

Detecting changes in the distribution of a stochastic process is a longstanding problem and a myriad of methods has been proposed; see e.g \cite{basse88:auto}. 
 Our concern is  the detection of non-Gaussian signals in a  \textit{Gaussian}  background, \textit{i.e.} \textit{Normality tests}. 
The framework considered here is one in which time-series are recorded on $d$ sensors (typically $d=2$ or $3$ in many applications). Moreover, we are interested in the \textit{online} problem of reacting to a change as quickly as possible after it occurs, also known as \textit{sequential detection} problem \cite{bass93:phec, poor08:cup}. 

Normality tests belong to the the class of tests without alternative, unlike other approaches based on the Likelihood Ratio   such as the CUSUM test \cite{page54:biom, hink71:biom}.
The case of i.i.d (scalar or $d$-variate) processes has received significant attention, see \cite{Henz02:statp}, \cite{Moor82:as} for a detailed survey. On the other hand, few tests concern the case of dealing with time-series from multiple sensors that in practical applications cannot be considered to be i.i.d, a case we will refer to as \textit{n.i.d} or \textit{colored}. 
Moreover, the proposed approaches for dealing with multivariate colored processes, such as \cite{Hini82:jtsa} \cite{MoulCC92:ssap},  
can hardly  be used in real-time applications. For this reason, we have derived a joint normality test for multivariate colored data \cite{Elbo21} that is implemented with low computational burden.

There has also been interest in testing the normality of \textit{unobserved} regression residuals \cite{jarq87:isc}, also referred to as the process of innovations. More precisely, define the Multi-dimensional Auto-Regressive \cite{lutk13:spri} (or Vector AR of order $p$ denoted VAR($p$)) model to describe the statistical behavior of the $d \times 1$ vector of observation $\vect{x}(i)$ for $i=1,\dots,N$:
\begin{equation}\label{eq:ar}
    \vect{x}(i) = \sum_{k=1}^{p}\matr{A}_{k}\vect{x}(i-k) + \vect{\epsilon}(i)
\end{equation}
where  $\matr{A}_{k}$ is a $d \times d$ matrix of unknown parameters and $\vect{\epsilon}(i)$ is the $i$th unobservable residual assumed zero-mean and i.i.d. An additional assumption  is that residuals are drawn from a normal distribution. 
The drawbacks of violating the latter assumption has been studied, for instance \cite{hogg79:tas} showed that the ordinary least squares method, which is usually used to estimate $\{\matr{A}_k\}_{1 \leq k\leq p}$, is sub-optimal for heavy-tailed distributions. Thus, it is important to validate this assumption of normality in  this  linear  model.

Moreover, since the residuals are estimated in practice, we expect that errors in the   model  specification and estimation will impact the whitening performance of the filter and the estimated innovation process $\widehat{\vect{\epsilon}}$ could no longer be considered i.i.d. 
In this case, the normality tests designed for i.i.d processes become biased as shown in \cite{Moor82:as},
highlighting the importance of deriving our joint normality test for variables that are not statistically independent. 

Within this  framework, we concentrate on the \emph{Multivariate Kurtosis} (MK) defined in (\ref{estimKurt-eq}). In \cite{Elbo21}, we calculated the power of this test variable in the colored case, and in \cite{elbo22:icassp} we studied its performances when the observed multivariate process was projected onto an arbitrary subspace of low dimension (typically $1$-$D$ or $2$-$D$), in particular for time series generated by colored copulas. 
The main contributions of the present paper are now the following:
\begin{itemize}
    \item We compare the performances of the MK test with and without linear prewhitening, \emph{i.e.} using $\vect{x}(n)$ or  $\vect{\epsilon}(n)$.
    \item We observe the impact of projecting $3$-$D$ observations and their \emph{regression residuals} onto an arbitrary plane ($2$-$D$ projection) or  direction ($1$-$D$ projection).
    \item For sake of time and memory effectiveness,  the test statistics and the regression function are estimated recursively,  by using both exponential averaging and the  Recursive Least Squares (RLS) algorithm. 
    \end{itemize}

The paper is organized as follows: our recent results \cite{Elbo21} are summarized in Section \ref{sec:test_statistic_definition}. The recursive implementation is described in Section \ref{sec:adaptive_ar}, and the complete algorithm in Section \ref{sec:change_detection}. Computer results are eventually reported in Section \ref{sec:experiments}.

\section{Asymptotic Statistics of The Kurtosis}\label{sec:test_statistic_definition}

Let $\vect{x}(n) =[x_{1}(n),x_{2}(n),\dots,x_{d}(n)]^{T} \in \RR^{d}$ be a real-valued $d$-variate random variable. Let $\matr{S}(\tau)=\E{\vect{x}(n) \vect{x}(n-\tau)^{T}}$ be the covariance function whose entries are $S_{ab}(\tau)$. Also denote $\matr{S}(0)=\matr{S}$. 

Let $\matr{X}\eqdef\{\vect{x}(1),\dots,\vect{x}(N)\}$ be a stochastic stationary process of $N$ random variables $\vect{x}(n)$. Our aim is to test:
\begin{equation}\label{eq-testGeneral}
    H_0: \matr{X} ~ \underset{n.i.d}{\sim} \mathcal{N}(0,\matr{S}) \quad versus \quad \bar{H}_0
\end{equation}
Where $\vect{x}(n)$ are identically but not independently distributed (n.i.d). 
Following Mardia's definition of the Multivariate Kurtosis (MK) \cite{Mard70:biom}, our test statistic reads:\vspace{-1ex}
\begin{equation}\label{estimKurt-eq}
     \hat{B}_d(N) = \frac{1}{N} \sum_{n=1}^{N}\big(\vect{x}(n)^T\widehat{\matr{S}}^{-1}\vect{x}(n)\big)^2 \vspace{-1ex}  
\end{equation}
with\vspace{-1ex} 
\begin{equation}
    \widehat{\matr{S}}=\frac{1}{N}\sum_{k=1}^{N}\vect{x}(k)\vect{x}(k)^{T}
\end{equation}

\subsection{Known results: Mardia's MK test for a i.i.d. process}
The following results make use of Landau's notations $o(\frac{1}{N})$ and $O(\frac{1}{N})$, to precise that the absolute approximation error is \textit{dominated by}  $\frac{1}{N}$ or is \emph{of the order of}  $\frac{1}{N}$,  respectively. 
\begin{theorem}\cite{Mard70:biom}\label{Mardiatest}
Let $\matr{X}$ be an \textit{i.i.d.} process of dimension $d$. If $\matr{X} ~ \underset{i.i.d}{\sim} \mathcal{N}(0,\matr{S})$ then 
$\hat{B}_d(N) \xrightarrow[N \to \infty]{} \mathcal{N}(\mu, \sigma^2)$, with: \vspace{-1ex minus 1ex} 
\begin{eqnarray}\label{eq:mardia_test_stats}
\mu &=&  d(d+2)+ o(\frac{1}{N}) \\
\sigma^2 &=& \frac{8d(d+2)}{N}+o(\frac{1}{N}) 
\end{eqnarray}
\end{theorem}

In \cite{Elbo21} we devised a similar theorem for multivariate data that are no longer considered i.i.d. In the following, we sketch the main assumptions to identify the statistics of Mardia's MK  test in the multivariate n.i.d. case.  
\subsection{Joint normality test for a colored process}\label{subsec:math_prelim}
Proofs of the results presented in this subsection can be found in \cite{Elbo21}, as well as theoretical details.

\subsubsection{Mathematical Preliminaries}
In order to guarantee convergence while relaxing the i.i.d assumption, we make the mixing assumption: 
$\sum_{\tau=0}^{\infty} |S_{ab}(\tau)|^2$  converges to a finite limit $\forall (a,b)\in\{1,\dots,d\}^2$. 

Let $\widehat{\matr{S}}=\matr{S}+\matr{\Delta}$, where $\matr{\Delta}$ is small compared to $\matr{S}$, then we have the following lemma:
\begin{lemma}\label{lemma1}
The entries of matrix $\matr{\Delta}$ are of order $O(1/\sqrt{N})$. 
\end{lemma}
 Let $\matr{G}=\matr{S}^{-1}$.  This lemma allows the following expansion of the precision matrix $\widehat{\matr{G}} \eqdef \widehat{\matr{S}}^{-1}$ up to second order in $\matr{\Delta}$:
\begin{equation}\label{eq:expansionG}
\widehat{\matr{G}} = \matr{G} - \matr{G}\matr{\Delta}\matr{G} + 
\matr{G}\matr{\Delta}\matr{G}\matr{\Delta}\matr{G} + o(1/N). 
\end{equation}
Finally, by injecting (\ref{eq:expansionG}) in $\hat{B}_d(N)$ defined in (\ref{estimKurt-eq}), an approximation of the test statistic in $o(N^{-1})$ may be obtained, 
as $\hat{B}_d$ can be shown to converge to a normal variable. Thus we only need to calculate its mean and  variance, whose expressions are given below.

\subsubsection{Statistics for a colored scalar process}\label{StatB1nid}
\begin{equation}\label{meanb1}
    \E{\hat{B}_{1}} = 3 - \frac{6}{N} - \frac{12}{N^2} \sum_{\tau=1}^{N-1} (N-\tau) \,\frac{S(\tau)^2}{S^2}  + o(\frac{1}{N})
\end{equation}
\begin{equation}\label{varb1}
    \mathbb{V}\text{ar}\{\hat{B}_{1}\} = \frac{24}{N}\Big[1+\frac{2}{N}\sum_{\tau=1}^{N-1}(N-\tau)  \frac{S(\tau)^{4}}{S^{4}}\Big]   + o(\frac{1}{N}) 
\end{equation}
The dependence between time samples is taken into account in the covariance function terms $S(\tau)$. In the case of i.i.d samples, we recover the expressions given in Theorem \ref{Mardiatest} above, for $d=1$.

\subsubsection{Statistics for a bivariate colored process}
\begin{equation}\label{meanb2}
    \E{\hat{B}_{2}} = 8 - \frac{16}{N} -  \frac{4}{N^2}\sum_{\tau=1}^{N-1} \frac{(N-\tau) Q_{1}(\tau)}{(S_{11}S_{22}-S_{12}^2)^2} + o(\frac{1}{N}) 
\end{equation}

\begin{equation}\label{varb2}
    \mathbb{V}\text{ar}\{\hat{B}_{2}\}=\frac{64}{N}+\frac{16}{N^{2}}\sum_{\tau=1}^{N-1}\frac{(N-\tau)Q_{2}(\tau)}{(S_{11}S_{22}-S_{12}^{2})^4} + o(\frac{1}{N}) 
\end{equation}
 where $Q_i(\tau)$ are linear combinations of $S_{ab}(\tau)$ thus containing information about the time-dependence between samples.  Despite  the difficulty to derive such expressions, the computational burden involved by their implementation is low. 


\section{Times Series Model: Multidimensional Auto-regressive Model}\label{sec:adaptive_ar}

\subsection{Definition}\label{subsec:varying_decay_factor}
The VAR($p$) model defined in (\ref{eq:ar}) can be written in matrix format as:
\begin{equation}\label{VAR_matrix_format}
    \matr{Y} = \matr{Z}\matr{w} + \matr{\epsilon},
\end{equation}
where, $\forall i\in \{p,\ldots,N-1\}$: 
\begin{eqnarray}
    \matr{w} &=& \big[ \matr{A}_1, \matr{A}_2, \hdots, \matr{A}_p \big]^T \in \mathbb{R}^{dp\times d} \label{weightmatrix}\nonumber\\
   \matr{Y} &=& \big[\vect{x}(p+1),\dots,\vect{x}(N) \big]^T \in   \mathbb{R}^{(N-p)\times d} \label{target}\nonumber\\ \fin
\matr{Z} &=& \big[ \vect{z}(p),\vect{z}(p+1),\hdots,\vect{z}(N-1)  \big]^T \in \mathbb{R}^{(N-p) \times dp} \label{designmatrix}\nonumber \\
    \vect{z}(i) &=& \big[ \vect{x}(i);\vect{x}(i-1);\hdots;\vect{x}(i-p+1) \big] \in \mathbb{R}^{dp \times 1}  \nonumber\\
\matr{\epsilon} &=& \big[\vect{\epsilon}(p+1),\vect{\epsilon}(p+2), \hdots, \vect{\epsilon}(N) \big]^T \in \RR^{(N-p) \times d} \label{vectorinnovations}\nonumber \fin
\end{eqnarray}
A Least Squares estimation can be performed to find $\matr{\hat{w}}$ that minimizes $\| \matr{Z}\matr{w} - \matr{Y} \|_2$ assuming that $\matr{\Gamma}=\matr{Z}^T \matr{Z}$ is non-singular:\vspace{-1ex minus 1ex} 
\begin{equation}\label{Multi_LS_estimator}
    \matr{\hat{w}} = \matr{\Gamma}^{-1}\matr{Z}^T\matr{Y}
\end{equation}

\subsection{Recursive Estimation of VAR($p$) parameters}
To estimate the regression parameters, a classical Recursive Least Squares methods (RLS) is proposed. This adaptive strategy is well known to be suitable for data showing  some sort of non-stationarity, and has been widely used (see e.g.  \cite{plackett1950some} \cite{rao70:jrss}, \cite{hall78:jma} and \cite{gren83:ieee} to only cite few). 

Below, the main RLS equations are given for the present problem. Let  $\lambda_1$ be the decay factor weighting the observations, and  
${\vect{\hat w}_k}$ be the $k^{\textnormal{th}}$ column of $\widehat{\matr{w}}$ defined in (\ref{weightmatrix}), and $\vect{y}_k(t-1) =\big(x_{k}(1), x_{k}(2),\hdots, x_{k}(t-1)\big)^T$.
\begin{eqnarray}
\vect{y}_k(t-1) &=& \matr{Z}\vect{\hat w}_k(t-1) + \matr{\epsilon}_k(t-1) \\
\vect{\hat{w}_k}(t-1) &=& \matr{\Gamma}^{-1}(t-1)\matr{Z}^T\vect{y}_k(t-1)
\end{eqnarray}

Suppose we want to update the model with new observations $\vect{x}(t)$;  a new row $\vect{z}(t)^T$ is appended to $\matr{Z}$ in (\ref{designmatrix}), and a new observation $x_k(t)$ is  appended to $\vect{y}_k$. 
In the RLS algorithm, $\matr{\Gamma}^{-1}(t)$ and   $\vect{\hat{w}}_k(t)$ are recursively expressed for $t>N$ as
%

\begin{eqnarray}\label{inversion_lemma}
    \matr{\Gamma}^{-1}(t) &=&\Big(\lambda_1 \matr{\Gamma}(t-1) + \vect{z}(t)\vect{z}(t)^T \Big)^{-1}  \\
     &=& \lambda_1^{-1}\matr{\Gamma}^{-1}(t-1) - b\vect{u}\vect{u}^T 
\end{eqnarray} 
~\vspace{-3ex minus 0.5ex}
\begin{eqnarray}\label{eq:tvar_update}
    &\vect{\hat{w}}_k(t) = \matr{\Gamma}^{-1}(t)\matr{\tilde{Z}}^T\vect{\tilde{y}}_k \\
    &\!\!\!=\vect{\hat{w}}_k(t-1) -b (\vect{z}^T(\vect{\hat{w}}_k(t-1)+x_{k}(t)\vect{u}))\vect{u}\fin + x_{k}(t)\vect{u},& \nonumber 
\end{eqnarray} 
%
 $\vect{u}=\lambda_1^{-1}\matr{\Gamma}^{-1}(t-1)\vect{z}(t)$ and $b=(1+\vect{z}(t)^T \vect{u})^{-1}$. 
 Note that $\matr{\tilde{Z}}$ is $\matr{Z}$ in (\ref{designmatrix}) augmented with the row $\vect{z}(t)^T$, and $\vect{\tilde{y}}_k = \big[ \vect{y}_k(t-1); x_{k}(t)\big]$.
Note that we can also vary $\lambda_1$ and the order $p$ of the model to allow for more flexibility and adapt to local non-stationarities of the data, see e.g \cite{giur10:spe}, \cite{cho90:icassp}. This is not detailed in this communication. 
\section{Change Detection Algorithm}\label{sec:change_detection}
\subsection{The test in practice}
First we choose the nominal level of the test:
\begin{equation*}
    \alpha=\mathbb{P}(\text{choose }\bar{H}_0| H_0 \text{ is true}).
\end{equation*}
\begin{itemize}
    \item It has been shown in previous sections
    that $z=(\widehat{B}_d-\E{\widehat{B}_d})/\sqrt{\mathbb{V}ar\{\widehat{B}_{d}\}}$ is asymptotically normal.
    \item We reject the null hypothesis $H_0$ at a significance level $\alpha$ if:
    \begin{equation*}
    2(1-\Phi(z)) < \alpha
    \end{equation*}
    where $\Phi$ denotes the cumulative distribution function (cdf) of $\mathcal{N}(0,1)$. 
    \item In practice, the time-structure is unknown and the covariance (matrix) function entries $S_{ab}(\tau)$  are replaced by their sample counterparts at time $t$.
%
\end{itemize}
\subsection{Recursive normality test on regression residuals}\label{subsec:changealgo}
The proposed test statistic is computationally efficient i.e it can be easily computed over a sliding window or by using the exponential weighting technique 
to test the normality of the estimated regression residuals available at time $t$ assumed to follow a Gaussian distribution $\mathcal{N}(0, \matr{V}(t))$. Let $0<\lambda_{2} <1$:
\begin{eqnarray}
\matr{V}(t)\! &\!=\!&\!\lambda_1\matr{V}(t-1) + (1-\lambda_{1})\hat{\vect{\epsilon}}(t)\hat{\vect{\epsilon}}(t)^{T} \\
    \hat{B}_d(t)\! &\!=\!& \!\lambda_2 \hat{B}_d(t-1) + (1-\lambda_2) (\hat{\vect{\epsilon}}(t)^T \widehat{\matr{V}}^{-1}\!(t)\hat{\vect{\epsilon}}(t))^2. ~~~ 
\end{eqnarray}

The algorithm for sequentially detecting changes reads:
\begin{algorithmic}[hbt!]
\Require{$p \geq 1$, $0 <\lambda_1, \lambda_2 <1$, $\alpha$, $\delta$}
\State \textbf{Initialization:}
$\matr{\Gamma}(p)\gets \delta \matr{I}_{dp}$, $\hat{B}_d(p)\gets 0$, $\matr{V}(p)\gets \matr{I}_d$

\For{$p+1 \leq t \leq N$}
    \State \text{Update} $\matr{\Gamma}^{-1}(t)$\hspace{5mm}\Comment{using (\ref{inversion_lemma})}
    
    \State \text{Update} $\{ \widehat{\matr{A}}_{k}(t)\}_{1 \leq k \leq p}$ \hspace{1cm}\Comment{using (\ref{eq:tvar_update})}
    
    \State \text{Compute }$\hat{\vect{\epsilon}}(t) = \vect{x}(t) - \displaystyle \sum_{k=1}^{p}\widehat{\matr{A}}_{k} \vect{x}(t-k)$

    \State \text{Compute }$z = (\hat{B}_d(t) - \E{\hat{B}_d(t)})/\sqrt{\mathbb{V}ar(\hat{B}_d(t))} $
    
    \If {$2(1-\Phi(z))<\alpha$}
    \State Change is detected
    
    \Else \State No change
    \EndIf
\EndFor
\end{algorithmic}


Forgetting factors $\lambda_1$ and $\lambda_2$ (for 2nd and 4th order statistics respectively) are usually chosen by a rule of thumb, depending on the time-scale of the change. To give a better intuition of this factor, one can calculate the length $N_{\text{eff}}$ of a uniform sliding window that would yield the same estimator variance, when data are i.i.d. normal. This  leads  to $\lambda \propto 1/N_{\text{eff}}$. More details will be provided in a subsequent paper. 

For the RLS algorithm, a common choice is $\lambda_1 = 0.99$ which corresponds to using a virtual window's size $N_{\text{eff}}=\frac{2}{1-\lambda_1}$. 
As for the recursive estimation of the Kurtosis, to prevent the problem of slow convergence, we choose $N_{\text{eff}}= 1000$.

\section{Computer Results}\label{sec:experiments}
A set of  of  Monte Carlo simulations is presented  to compare the power of the proposed normality test  when applied directly on data or on regression residuals. Then, we study the performance of the test statistic   on a low-dimensional ($2$-$D$ or $1$-$D$) projection of the initial multivariate data.  As a final illustration of the effectiveness of  our  method, we apply the change detection Algorithm presented in subsection \ref{subsec:changealgo} on synthetic colored data undergoing an abrupt change in its distribution.

\subsection{Applying the test on colored data}

\subsubsection{Directly on data}\label{subsec:exp_protocole}  $M=2000$ simulations are considered, each being based on  a sequence of $N=1000$ samples. First, we simulate 1-D  AR($p$) processes (for $p \in \{4, 14, 20 \})$. The AR coefficients are computed such that the equivalent filter is low-pass, with band pass equal to .25 (normalized freq).



For each  simulation a 2-D Gaussian (or Uniform)  AR($p$) process is constructed by time embedding : $\vect{x}(t)=\{x(2t),x(2t+1)\}$. Then, both Mardia's test derived for i.i.d. samples (denoted $\hat{B}_{1,i.i.d}$) and the test for colored samples, whose statistics are defined by equations (\ref{meanb1}, \ref{varb1}) for colored samples (denoted  $\hat{B}_1$)  are applied on the marginals of the 2-D process, and compared.  Finally, the statistics derived for n.i.d. bi-variate data $\hat{B}_{2}$,   described by equations (\ref{meanb2}, \ref{varb2}) is also applied  to the 2-D process. 

The obtained empirical rejection rates of hypothesis $H_0$ are computed as $\frac{\# Rejections}{M}$, for a  test level $\alpha= 5\%$. The results are summarized in Tables \ref{tab:direclty_on_data} and \ref{tab:direclty_on_data_ar20}.

\begin{table}[htbp]
\renewcommand{\arraystretch}{1.3}
\centering
\caption{Empirical Rejection Rates for $2000$ simulations for $\alpha=5\%$ significance level with the $\hat{B}_{1,i.i.d}, \hat{B}_1, \hat{B}_2$ test applied directly on AR($p$) data with $p=4, 14$}
\begin{tabular}{l|ll|ll|}
\cline{2-5}
                                        & \multicolumn{2}{c|}{AR($4$)}            & \multicolumn{2}{c|}{AR($14$)}       \\\hline
\multicolumn{1}{|l|}{Test Statistic}    & \multicolumn{1}{l|}{Gaussian} & Uniform & \multicolumn{1}{l|}{Gaussian} & Uniform                      \\ \hline
\multicolumn{1}{|l|}{$\hat{B}_{1,iid}$} & \multicolumn{1}{l|}{$0.067$}  & $1.$    & \multicolumn{1}{l|}{$0.123$}  & $0.512$                          \\
\multicolumn{1}{|l|}{$\hat{B}_{1}$}     & \multicolumn{1}{l|}{$0.052$}  & $0.99$  & \multicolumn{1}{l|}{$0.045$}  & $0.456$                            \\
\multicolumn{1}{|l|}{$\hat{B}_{2}$}     & \multicolumn{1}{l|}{$0.06$}   & $1.$    & \multicolumn{1}{l|}{$0.065$}  & $0.88$                             \\ \hline
\end{tabular}
\label{tab:direclty_on_data}
\end{table}

\begin{table}[htbp]
\renewcommand{\arraystretch}{1.1}
\centering
\caption{Empirical Rejection Rates for $2000$ simulations for $\alpha=5\%$ significance level with the $\hat{B}_{1,i.i.d}, \hat{B}_1, \hat{B}_2$ test applied directly on AR($20$) process}
\begin{tabular}{l|ll|}
\cline{2-3}
                                        & \multicolumn{2}{c|}{AR($20$)}          \\ \hline
\multicolumn{1}{|l|}{Test Statistic}    & \multicolumn{1}{l|}{Gaussian} & Uniform  \\ \hline
\multicolumn{1}{|l|}{$\hat{B}_{1,iid}$} & \multicolumn{1}{l|}{$0.228$}  & $0.430$                       \\
\multicolumn{1}{|l|}{$\hat{B}_{1}$}     & \multicolumn{1}{l|}{$0.047$}  & $0.399$                     \\
\multicolumn{1}{|l|}{$\hat{B}_{2}$}     & \multicolumn{1}{l|}{$0.06$}   & $0.688$ \\ \hline
\end{tabular}
\label{tab:direclty_on_data_ar20}
\end{table}

\begin{itemize}
    \item It seems that $\hat{B}_{1,i.i.d}$ has a better detection power than $\hat{B}_1$. As a matter of fact, by comparing the formulas of their variance in (\ref{eq:mardia_test_stats}, \ref{varb1}), we can see that that the variance of $\hat{B}_{1,i.i.d}$ is underestimated, consequently, the latter over-rejects the hypothesis of Gaussianity. This is noticeable even for Gaussian AR($p$) process (Under $H_0$), with rejection rates that surpass the nominal level $5 \%$.
    \item Both scalar tests $\hat{B}_{1,i.i.d}$ and $\hat{B}_1$ perform poorly compared to the joint normality test statistic $\hat{B}_2$.
    \item When the correlation tails last longer (Table \ref{tab:direclty_on_data_ar20}), the overall performance of the test statistics tends to decrease. 
\end{itemize}

\subsubsection{On Regression residuals} We now study the power of the normality tests on \emph{estimated} regression residuals. We utilize the Least Squares method to obtain $\hat{\epsilon}$. The simulation procedure and the tests are the same as those described in the previous paragraph \ref{subsec:exp_protocole}. We study the case where the order $p$ of the generated AR($p$) process is known (we choose $p=20$), and the case where the order is misspecified ($\hat{p}=9$). The empirical rejection rates are summarized in Table \ref{tab:on_regression_residuals}.

\begin{table}[htb]
\renewcommand{\arraystretch}{1.3}
\caption{Empirical Rejection Rates for $2000$ simulations for $\alpha=5\%$ significance level with the test statistics $\hat{B}_{1,i.i.d}, \hat{B}_1, \hat{B}_2$ applied on estimated regression residuals using OLS method, with known and misspecified order}
\begin{tabular}{l|ll|ll|}
\cline{2-5}
 & \multicolumn{2}{c|}{On Residuals of AR($20$)} &
   \multicolumn{2}{c|}{On Residuals of AR($20$)}\\ 
 & \multicolumn{2}{c|}{$p=20$} & 
   \multicolumn{2}{c|}{$\hat{p}=9$} \\ 
\hline
\multicolumn{1}{|l|}{Test Statistic}    & \multicolumn{1}{l|}{Gaussian} & Uniform &  \multicolumn{1}{l|}{Gaussian}                         & Uniform                        \\ \hline
\multicolumn{1}{|l|}{$\hat{B}_{1,iid}$} &  \multicolumn{1}{l|}{$0.06$}                           & $1.$                            & \multicolumn{1}{l|}{$0.064$}                          & $0.582$                        \\
\multicolumn{1}{|l|}{$\hat{B}_{1}$}     &  \multicolumn{1}{l|}{$0.05$}                           & $1.$                            & \multicolumn{1}{|l|}{$0.051$}                          & $0.429$                        \\
\multicolumn{1}{|l|}{$\hat{B}_{2}$}     & \multicolumn{1}{l|}{$0.055$}                          & $1.$                            & \multicolumn{1}{|l|}{$0.06$}                           & $0.850$                        \\ \hline
\end{tabular}
\label{tab:on_regression_residuals}
\end{table}

\begin{itemize}
    \item If the model is perfectly known, then all test statistics perform well on the well estimated regression residuals. However,  if the model order is under estimated, some important time correlation remain, and  scalar tests perform poorly compared to the joint normality test, as expected.  
\end{itemize}

\subsection{Random Projections}
We simulate a $3$-$D$ process  VAR($p$), $p \in \{ 5,20\}$,  of length $N=1000$ following equation (\ref{eq:ar}), 
where  the inputs \fin $\vect{\epsilon}(t)$ are i.i.d with distributions either multivariate standard normal $\mathcal{N}(0,1)$ or multivariate $\mathcal{U}(-2,2)$.
 
Then $M=2000$ different projections on an arbitrary plane ($2$-$D$ projection)  going through the origin of the 3-D space are computed, corresponding to as many 2-D time series. 
For comparison, the same set of observations is also projected $M$ times on an arbitrary direction ($1$-$D$ projection). 
Eventually, we run the same set of experiments on estimated regression residuals   estimated by ordinary least squares method (OLS). The results are reported below. 

\begin{table}[!htb]
\renewcommand{\arraystretch}{1.1}
\caption{Empirical rejection rates of the test statistics applied to a low representation of $3$-$D$ VAR($p$) process}
 \begin{minipage}{.49\linewidth}
 \centering 
   \begin{tabular}{l|ll|}
   \cline{2-3}                                                                & \multicolumn{2}{c|}{\begin{tabular}[c]{@{}c@{}}$3$-$D$ \\ VAR$(5)$\end{tabular}} \\ \hline
   \multicolumn{1}{|l|}{\begin{tabular}[c]{@{}l@{}} $2$-$D$ \\ projection\end{tabular}} & \multicolumn{1}{l|}{Gaussian}                     & Uniform                    \\ \hline
   \multicolumn{1}{|l|}{$\hat{B}_{2}$}                                                                   & \multicolumn{1}{l|}{$0.051$}                      & $0.986$                    \\    
   \hline
   \end{tabular}
 \end{minipage}\hspace{1.4mm}%
 \begin{minipage}{.49\linewidth}
 \centering 
   \begin{tabular}{l|c|c|}
   \cline{2-3}                                                                          & \multicolumn{2}{c|}{\begin{tabular}[c|]{@{}c@{}}$3$-$D$ \\ VAR$(5)$\end{tabular}} \\ 
   \hline
   \multicolumn{1}{|l|}{\begin{tabular}[c]{@{}l@{}} $1$-$D$ \\ projection\end{tabular}} & \multicolumn{1}{l|}{Gaussian}                      & Uniform                     \\ 
   \hline
   \multicolumn{1}{|l|}{$\hat{B}_{1}$}                                                  & \multicolumn{1}{l|}{$0.056$}                       & $0.529$   \\ 
   \hline
   \end{tabular}
 \end{minipage} 
\end{table}

\begin{table}[htb]
\renewcommand{\arraystretch}{1.3}
\centering
\caption{Empirical rejection rates of the test statistics applied to a low representation of $3$-$D$ VAR($20$) process  with uniform inputs and its estimated regression residuals with a VAR($10$) model}
\begin{tabular}{l|c|c|}
\cline{2-3}
    & $3$-$D$   & On residuals\\  
    & VAR$(20)$ & $\hat{p}=10$      \\ 
\hline
\multicolumn{1}{|l|}{$1$-$D$  proj., $\hat{B}_{1}$}  & $0.250$                                                      & $0.41$                                                          \\ \hline
\multicolumn{1}{|l|}{$2$-$D$  proj.,  $\hat{B}_{2}$} & $0.580$                                                      & $0.9$                                                          \\ \hline
\end{tabular}
\label{tab:proj_on_residuals}
\end{table}

\begin{itemize}
    \item  The test $\hat{B}_1$ applied directly on 1-D projections of either the observations or its residuals computed by OLS, performs poorly. This is in accordance with our observations from the preceding experiments.  
    \item Even with a misspecified order, the joint test statistic performs best (empirical power of $90\%$) on the $2$-$D$ low representation of regression residuals,  as it is able to account for both temporal and spatial (between coordinates) dependences.
\end{itemize}
\subsection{Change detection on synthetic data}
Consider the case where a process $\epsilon$ is constituted of i.i.d samples following a standard normal distribution. The process undergoes a change at $n_c = 5000$ in its distribution: samples in the interval $[5000, 10000]$ are now following a uniform distribution $\mathcal{U}(-\sqrt{3},\sqrt{3})$. The change ends at $n = 10000$ and the samples are again normally distributed. The process $\epsilon$ is then (low pass) filtered using an AR($5$)  model and is now denoted $x$. 

\begin{figure}[htb]
    \centering
    \includegraphics[width=.47\textwidth]{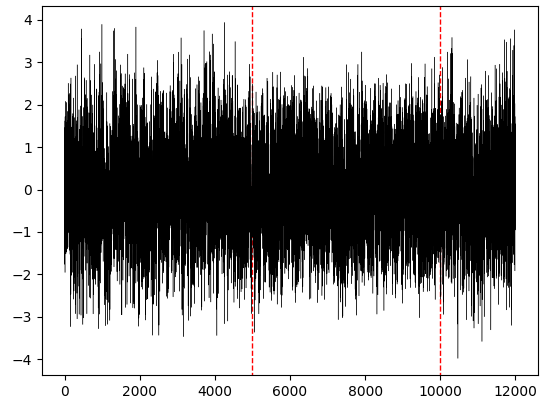}\vspace{-2ex minus 0.5ex}
    \caption{ One realization of a Gaussian AR($5$) process that undergoes an abrupt change in the distribution of its excitation $\epsilon$ (from $\mathcal{N}(0,1)$ to $\mathcal{U}(-\sqrt{3},\sqrt{3})$). Affected samples are between red dashed lines. }
    \label{fig:input_data_simu}
\end{figure}

One realization of this process is given in Fig. \ref{fig:input_data_simu}. It is clear from this figure that the change in the distribution is unbeknownst to the human eye.

The change detection algorithm presented in subsection \ref{subsec:changealgo} is applied to this realization by setting $p=5, \lambda_1 = 0.99, \lambda_2 = 0.998$,
$\alpha=5\%$ and $\delta=1$. A $2$-$D$ process is obtained by taking $\vect{x}(t)=\{x(2t), x(2t+1)\}$. 

For comparison, the CUSUM algorithm \cite{bass93:phec} is applied on the regression residuals by using its recursive form. The instantaneous log-likelihood ratio is computed as:
\begin{equation}\label{eq:LR}
    L(t) = -\ln(2\sqrt{3})+\frac{1}{2}\ln(2\pi)+\frac{1}{2} \hat{\epsilon}^2(t) \vspace{-3ex} 
\end{equation}

\begin{figure}[htb]
    \centering
    \includegraphics[width=.47\textwidth]{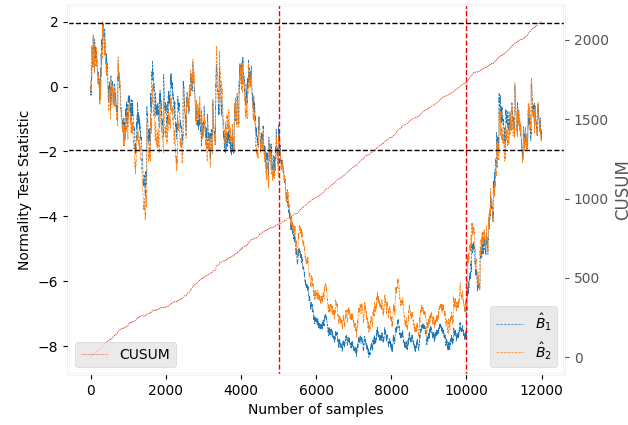}\vspace{-2ex minus 0.5ex}
    \caption{ Evolution of the normalized test statistics $\hat{B}_1$ (in blue) and $\hat{B}_2$ (in orange). In red, the evolution of the cumulative sum $s(n)=\sum_{t=1}^{n}L(t)$. Black horizontal dashed lines are the critical values $\pm 1.96$ corresponding to a test power of $95\%$. Red vertical dashed lines are the beginning and end  of an abrupt change  in the excitation statistics of an AR($5$) process.
    }
    \label{fig:simu_results}
\end{figure}
\fin

\newpage
\begin{itemize}
    \item Our proposed test statistics $\hat{B}_{1,2}$ stay in between $\pm 1.96$; in other words they do not reject the null hypothesis of Gaussianity with a false alarm rate of $5\%$.  They grow continuously in absolute value after the change time $n_c = 5000$ and keep rejecting $H_0$ until the end of the change at $n=10000$.
    \item There is no clear-cut in the cumulative sum algorithm that indicates a change in the distribution of the residuals. In fact, the likelihood ratio in (\ref{eq:LR}) is derived under the hypothesis that the process $\epsilon$ is i.i.d. As the latter's values are estimated recursively, they are more likely to have residual correlations between them and the hypothesis of \emph{independence} no longer holds, explaining why the cumulative sum of $L(t)$ keeps increasing.  
\end{itemize}

\section{Concluding remarks}

A novel approach to testing the Gaussianity of colored data is proposed. In a first stage, data  are whitened recursively. Then the MK test is applied in a second stage on available regression residuals in an \emph{online} manner. Computer experiments have been performed to assess the impact of the preprocessing stage on the power of the test. 
They have evidenced the fact that the bivariate MK statistics we recently obtained  (assuming both \emph{spatial} and \emph{temporal} dependence) perform always better than their  scalar counterparts. 
The latter are indeed severely affected by serial correlation in residuals due to imperfect whitening in the first stage. The same conclusions can be drawn if data are randomly projected onto a low-dimensional subspace. 

Future work will concern the validation of the change detection algorithm on real-world phenomena recorded by a network of $d$-variate sensors within a multiple hypothesis testing framework.

\bibliographystyle{./IEEEtran}
\bibliography{references}

\begin{thebibliography}{10}
\providecommand{\url}[1]{#1}
\csname url@samestyle\endcsname
\providecommand{\newblock}{\relax}
\providecommand{\bibinfo}[2]{#2}
\providecommand{\BIBentrySTDinterwordspacing}{\spaceskip=0pt\relax}
\providecommand{\BIBentryALTinterwordstretchfactor}{4}
\providecommand{\BIBentryALTinterwordspacing}{\spaceskip=\fontdimen2\font plus
\BIBentryALTinterwordstretchfactor\fontdimen3\font minus
  \fontdimen4\font\relax}
\providecommand{\BIBforeignlanguage}[2]{{%
\expandafter\ifx\csname l@#1\endcsname\relax
\typeout{** WARNING: IEEEtran.bst: No hyphenation pattern has been}%
\typeout{** loaded for the language `#1'. Using the pattern for}%
\typeout{** the default language instead.}%
\else
\language=\csname l@#1\endcsname
\fi
#2}}
\providecommand{\BIBdecl}{\relax}
\BIBdecl

\bibitem{basse88:auto}
M.~Basseville, ``Detecting changes in signals and systems—a survey,''
  \emph{Automatica}, vol.~24, no.~3, pp. 309--326, 1988.

\bibitem{bass93:phec}
M.~Basseville, I.~V. Nikiforov \emph{et~al.}, \emph{Detection of abrupt
  changes: theory and application}.\hskip 1em plus 0.5em minus 0.4em\relax
  prentice Hall Englewood Cliffs, 1993, vol. 104.

\bibitem{poor08:cup}
H.~V. Poor and O.~Hadjiliadis, \emph{Quickest detection}.\hskip 1em plus 0.5em
  minus 0.4em\relax Cambridge University Press, 2008.

\bibitem{page54:biom}
E.~S. Page, ``Continuous inspection schemes,'' \emph{Biometrika}, vol.~41, no.
  1/2, pp. 100--115, 1954.

\bibitem{hink71:biom}
D.~V. Hinkley, ``Inference about the change-point from cumulative sum tests,''
  \emph{Biometrika}, vol.~58, no.~3, pp. 509--523, 1971.

\bibitem{Henz02:statp}
R.~Henze, ``Invariant tests for multivariate normality: a critical review,''
  \emph{Statistical papers}, vol.~43, pp. 467--506, 2002.

\bibitem{Moor82:as}
D.~S. Moore, ``The effect of dependence on chi squared tests of fit,''
  \emph{The Annals of Statistics}, vol.~10, no.~4, pp. 1163--1171, 1982.

\bibitem{Hini82:jtsa}
M.~Hinich, ``Testing for {G}aussianity and linearity of a stationary time
  series,'' \emph{Jour. Time Series Analysis}, vol.~3, no.~3, pp. 169--176,
  1982.

\bibitem{MoulCC92:ssap}
E.~Moulines, K.~Choukri, and M.~Charbit, ``Testing that a multivariate
  stationary time series is {G}aussian,'' in \emph{Sixth {SSAP} Workshop on
  Stat. Signal and Array Proc.}, Oct. 1992, pp. 185--188.

\bibitem{Elbo21}
\BIBentryALTinterwordspacing
S.~ElBouch, O.~Michel, and P.~Comon, \emph{{A Normality Test for Multivariate
  Dependent Samples}}, Sep. 2021, hal-03344745. [Online]. Available:
  \url{https://hal.archives-ouvertes.fr/hal-03344745}
\BIBentrySTDinterwordspacing

\bibitem{jarq87:isc}
C.~M. Jarque and A.~K. Bera, ``A test for normality of observations and
  regression residuals,'' \emph{Int. Statistical Review}, pp. 163--172, 1987.

\bibitem{lutk13:spri}
H.~L{\"u}tkepohl, \emph{Introduction to multiple time series analysis}.\hskip
  1em plus 0.5em minus 0.4em\relax Springer Science \& Business Media, 2013.

\bibitem{hogg79:tas}
R.~V. Hogg, ``Statistical robustness: One view of its use in applications
  today,'' \emph{The American Statistician}, vol.~33, no.~3, pp. 108--115,
  1979.

\bibitem{elbo22:icassp}
\BIBentryALTinterwordspacing
S.~ElBouch, O.~Michel, and P.~Comon, ``{Joint Normality Test Via
  Two-Dimensional Projection},'' in \emph{ICASSP}, Singapore, May 2022.
  [Online]. Available: \url{https://hal.archives-ouvertes.fr/hal-03369151}
\BIBentrySTDinterwordspacing

\bibitem{Mard70:biom}
K.~V. Mardia, ``Measures of multivariate skewness and kurtosis with
  applications,'' \emph{Biometrika}, vol.~57, pp. 519--530, 1970.

\bibitem{plackett1950some}
R.~L. Plackett, ``Some theorems in least squares,'' \emph{Biometrika}, vol.~37,
  no. 1/2, pp. 149--157, 1950.

\bibitem{rao70:jrss}
T.~S. Rao, ``The fitting of non-stationary time-series models with
  time-dependent parameters,'' \emph{Journal of the Royal Statistical Society:
  Series B (Methodological)}, vol.~32, no.~2, pp. 312--322, 1970.

\bibitem{hall78:jma}
M.~Hallin, ``Mixed autoregressive-moving average multivariate processes with
  time-dependent coefficients,'' \emph{Journal of Multivariate Analysis},
  vol.~8, no.~4, pp. 567--572, 1978.

\bibitem{gren83:ieee}
Y.~Grenier, ``Time-dependent arma modeling of nonstationary signals,''
  \emph{IEEE Transactions on Acoustics, Speech, and Signal Processing},
  vol.~31, no.~4, pp. 899--911, 1983.

\bibitem{giur10:spe}
C.~D. Giurc{\u{a}}neanu and S.~A. Razavi, ``{AR} order selection in the case
  when the model parameters are estimated by forgetting factor least-squares
  algorithms,'' \emph{Signal Processing}, vol.~90, no.~2, pp. 451--466, 2010.

\bibitem{cho90:icassp}
Y.~Cho, S.~B. Kim, and E.~J. Powers, ``Time-frequency analysis using ar models
  with variable forgetting factors,'' in \emph{ICASSP}, 1990, pp. 2479--2482.

\end{thebibliography}

\end{document}